\documentclass[preprint,figures]{epl}
\usepackage{amsmath}
\usepackage{amssymb}

\usepackage{bm}

\begin{document}

\title{Novel properties of the Kohn-Sham exchange potential for open systems:
application to the two-dimensional electron gas}
\shorttitle{Novel properties of the Kohn-Sham exchange potential}

\author{S. Rigamonti\inst{1} \and C. R. Proetto\inst{1}\thanks{E-mail: \email{proetto@cab.cnea.gov.ar}} \and F. A.
Reboredo\inst{2}}
\shortauthor{S. Rigamonti \etal}

\institute{
\inst{1}Centro At\'{o}mico Bariloche and Instituto Balseiro - 8400
S. C. de Bariloche, R\'{\i}o Negro, Argentina\\
\inst{2}Lawrence Livermore National Laboratory - Livermore, CA 94551,
USA
}


\maketitle

\begin{abstract}
The properties of the Kohn-Sham (KS) exchange potential for open systems 
in thermodynamical equilibrium,
where the number of particles is non-conserved, are analyzed with
the Optimized Effective Potential (OEP) method of Density Functional
Theory (DFT) at zero temperature. The quasi two-dimensional electron
gas (2DEG) is used as an illustrative example. The main findings are
that the KS exchange potential builds a significant barrier-like structure
under slight population of the second subband, and that both the asymptotic
value of the KS exchange potential and the inter-subband energy jump
discontinuously at the one-subband (1S) $\rightarrow$ two-subband
(2S) transition. The results obtained in this system offer new insights
on open problems of semiconductors, such as the band-gap underestimation
and the band-gap renormalization by photo-excited carriers. 
\end{abstract}

Density Functional Theory (DFT) has become a most used computational
tool in the study of strongly inhomogeneous interacting systems such
as atoms, molecules, clusters, and solids.\cite{parr} A crucial step
for any DFT calculation is the approximation used to evaluate the
exchange-correlation $(xc)$ contribution $E_{xc}\equiv E_{x}+E_{c}$
to the total energy. Traditionally, $E_{xc}$ has been evaluated by
using density-dependent $xc$-functionals, either of local (LDA) or
semi-local (GGA, meta-GGA,...) character.\cite{kurth} In the last few
years, however, considerable interest has arisen around orbital-dependent
$xc$ functionals, which are implicit functionals of the density and
whose implementation in the KS scheme is known as the OEP method.\cite{grabo}
If the Hartree-Fock expression for the exchange energy functional
is used, and $E_{c}$ is neglected, the OEP method is equivalent to
the exact $x$-only implementation of KS theory.\cite{sahni} Several
advantages are associated to the use of exact $x$-only DFT calculations
for (closed) systems with a fixed number of particles: cancellation
of the spurious Hartree self-interaction energy,\cite{zunger} correct
high-density limit,\cite{kurth} great improvement in the KS eigenvalue
spectrum,\cite{dellasala} semiconductor band structure and excitations,\cite{semi}
and nonlinear optical properties.\cite{van} It is the aim of this
work to report on further novel features of the OEP scheme for open
configurations where the system under study is in thermodynamical equilibrium 
with a reservoir and
can exchange particles
with it, using a quasi-two-dimensional electron gas at zero
temperature to illustrate our results. We compare usual approximations
against our exact (OEP) $x$-only results.

\begin{figure}
\onefigure[  bb=2.4cm 13.5cm 19cm 26.2cm,
  clip,
  scale=0.5] {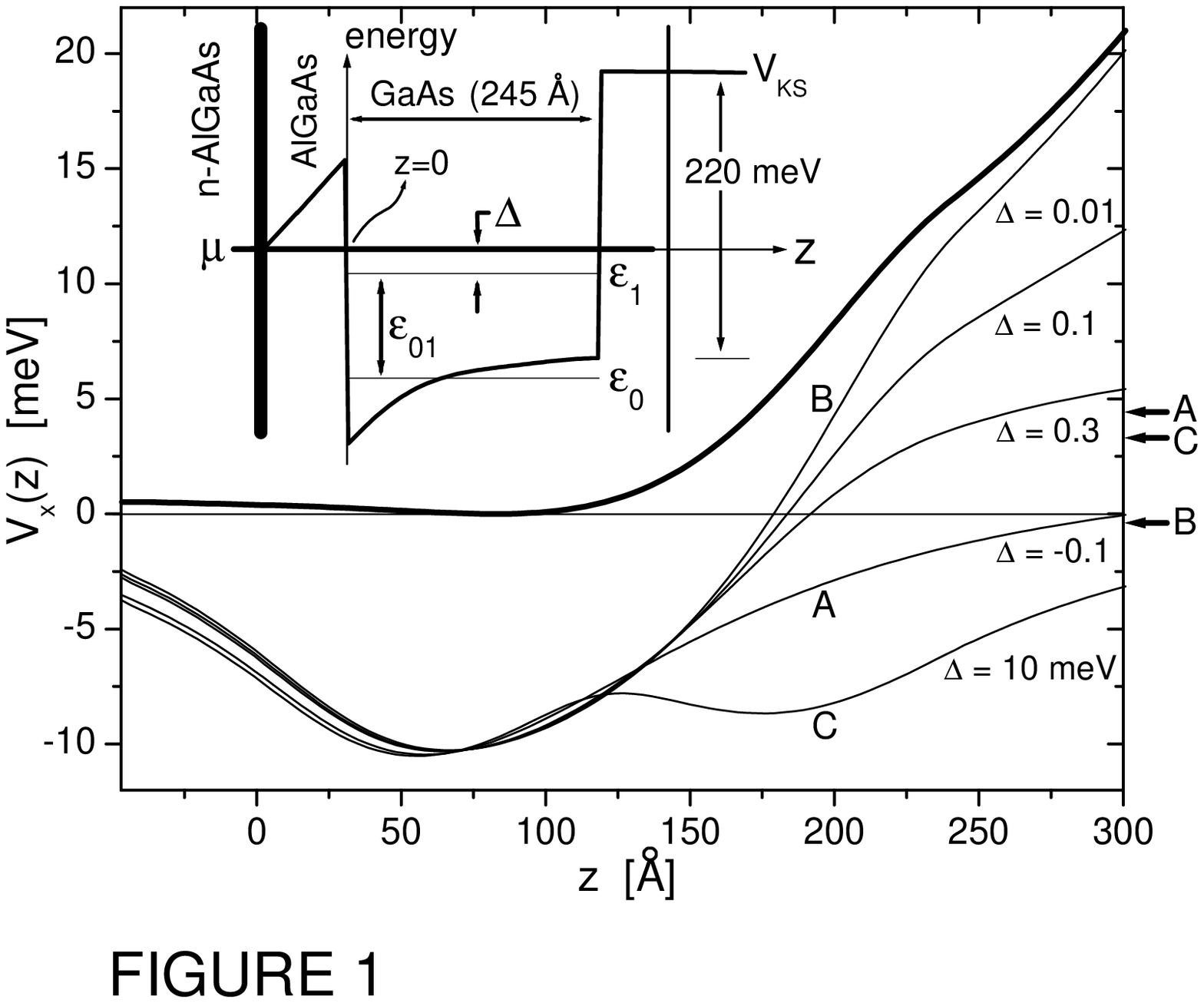}
\caption{Thin lines: exchange-only OEP for $\Delta=-0.1(A),0.01(B),0.1,0.3$
and $10$ meV $(C)$. Thick line: $V_{x\,2}(z)$ in the limit $k_{F}^{1}\rightarrow0^{+}$
(see text below). The arrows on the right denote the asymptotic
value of the exchange potential for selected subband fillings. Inset:
schematic view of our model for the modulation-doped QW. The thick
vertical stripe on the left represents the ionized donor impurities
region. The thin vertical line on the right represents a distant metallic
plane which induces  a charge-transfer field along $z$ included in $V_{ext}(z)$.
If the metallic plate is positively (negatively) charged more (less) electrons are transferred 
towards the well.}
\label{fig1}
\end{figure}

The present theory can be applied to any system with translational
symmetry in a plane. For instance, a semiconductor modulation-doped
quantum well (QW) grown epitaxially as shown in the inset of Fig.
\ref{fig1}. In these systems, it is possible to confine an electron gas by
changing the semiconductor in the growth direction $z.$ If the larger
gap semiconductor is doped with donors, it provides electrons to the
trap formed by the smaller gap semiconductor, until the charge-transfer
field equilibrates the \textit{common} chemical potential $\mu$ in the QW
and a doped region acting as particle reservoir. The charge transferred
from the reservoir to the QW can be tuned by an external electric
field. Assuming translational symmetry of the 2DEG in the $x-y$ plane
(area $A$), and proposing accordingly a solution of the type $\phi_{i\mathbf{k}\sigma}(\bm{\rho},z)
=\exp(i\mathbf{k}\cdot\bm{\rho})\xi_{i}^{\sigma}(z)/\sqrt{A}$,
the ground-state electron density can be obtained by solving a set
of effective one-dimensional Kohn-Sham equations of the form
\begin{equation}\label{eq1}
\widehat{h}_{KS}^{i\sigma}(z)\,\xi_{i}^{\sigma}(z)
\equiv\left[-\frac{\partial^{2}}{2\partial z^{2}}+V_{KS}^{\sigma}(z)-\varepsilon_{i}^{\sigma}\right]\xi_{i}^{\sigma}(z)=0,
\end{equation}
where effective atomic units have been used. $\xi_{i}^{\sigma}(z)$
is the eigenfunction for electrons in subband $i$ $(=0,1,...),$
spin $\sigma$ $(=\uparrow,\downarrow),$ and eigenvalue $\varepsilon_{i}^{\sigma}$.
The local (multiplicative) Kohn-Sham potential $V_{KS}^{\sigma}(z)$
is the sum of several terms: $V_{KS}^{\sigma}(z)=V_{ext}(z)+V_{H}(z)+V_{xc}^{\sigma}(z)$.
$V_{ext}(z)$ is given by the sum of the epitaxial potential plus
an external electric field. $V_{H}(z)$ is the Hartree potential.
Within DFT, $V_{xc}^{\sigma}(z)=\delta E_{xc}/\delta n_{\sigma}(z)$.
In an open system, $E_{xc}\equiv E_{xc}[\{\varepsilon_{i}^{\sigma}\},\{\xi_{i}^{\sigma}(z)\}]$
is a functional of the set of $\varepsilon_{i}^{\sigma}$'s and $\xi_{i}^{\sigma}(z)$'s.
The zero-temperature 3D electron density is $n_{\sigma}(z)=
\sum_{\varepsilon_{i}^{\sigma}<\mu}(k_{F}^{i\sigma})^{2}\left|\xi_{i}^{\sigma}(z)\right|^{2}/4\pi$,
with $k_{F}^{i\sigma}=\sqrt{2(\mu-\varepsilon_{i}^{\sigma})}.$ After
some lengthy but standard manipulations of the OEP scheme,\cite{grabo,rigamonti}
the calculation of $V_{xc}^{\sigma}(z)$ for an open system can be
summarized in the following set of equations: 
\begin{gather}
\label{eq2}\sum_{i}^{occ.}S_{i\sigma}(z)=0,\\
\label{eq3}S_{i\sigma}(z)=(k_{F}^{i\sigma})^{2}\psi_{i}^{\sigma}(z)^{\ast}\xi_{i}^{\sigma}(z)-C_{xc}^{i\sigma}\left|\xi_{i}^{\sigma}(z)\right|^{2}+c.c.,\\
\label{eq4}\psi_{i}^{\sigma}(z)=
\sum_{j\neq i}\frac{\xi_{j}^{\sigma}(z)}{\varepsilon_{i}^{\sigma}-\varepsilon_{j}^{\sigma}}
\int dz^{\prime}\xi_{j}^{\sigma}(z^{\prime})^{\ast}\Delta V_{xc}^{i\sigma}(z^{\prime})\xi_{i}^{\sigma}(z^{\prime}).
\end{gather}
The sum in Eq.(\ref{eq2}) runs over the occupied subbands, 
$C_{xc}^{i\sigma}=\overline{V}_{xc}^{i\sigma}+(2\pi/A)\partial E_{xc}/\partial\varepsilon_{i}^{\sigma},$
$\Delta V_{xc}^{i\sigma}(z)=V_{xc}^{\sigma}(z)-u_{xc}^{i\sigma}(z),$
$u_{xc}^{i\sigma}(z)=4\pi/A(k_{F}^{i\sigma})^{2}\xi_{i}^{\sigma}(z)^{\ast}\delta E_{xc}/\delta\xi_{i}^{\sigma}(z),$
and mean values are defined as $\overline{O}^{i\sigma}=\int dz\xi_{i}^{\sigma}(z)^{\ast}O(z)\xi_{i}^{\sigma}(z).$
Eqs. (\ref{eq2}-\ref{eq4}) determine the local $V_{xc}^{\sigma}(z)$ corresponding
to an orbital and eigenvalue-dependent approximation for $E_{xc}.$
Eqs. (\ref{eq1}) and (\ref{eq2}) have to be solved self-consistently.

Some comments are necessary here: \textit{a)} Our treatment corresponds
to an open system, and this is the origin of the term $\overline{V}_{xc}^{i\sigma}$ in
 $C_{xc}^{i\sigma}$ on the r.h.s. of Eq.(\ref{eq3}). This term gives a
finite $(\neq0)$ contribution under the replacement $V_{xc}^{\sigma}(z)\rightarrow V_{xc}^{\sigma}(z)+\alpha^{\sigma}$,
with $\alpha^{\sigma}$ an arbitrary constant. As a consequence Eq.(\ref{eq2})
is \textit{not} invariant if $V_{xc}^{\sigma}(z)$ is shifted by a
constant, and $V_{xc}^{\sigma}(z)$ is \textit{strictly} determined.
For closed systems, this term is absent and the corresponding OEP
expression gives $V_{xc}^{\sigma}(z)$ up to an additive constant.\cite{grabo}
\textit{b)} Using Eq.(\ref{eq4}) it can be found that the $\psi_{i}^{\sigma}(z)$
(denoted as {}``shifts'') satisfy the orthogonality constraint $\int dz\xi_{i}^{\sigma}(z)^{\ast}\psi_{i}^{\sigma}(z)=0$.
Application of $\hat{h}_{KS}^{i\sigma}(z)$ to the $\psi_{i}^{\sigma}(z)$
in Eq. (\ref{eq4}) yields an inhomogeneous differential equation, which can
be considered as an alternative definition of the shifts.\cite{method,kummel}
\textit{c)} Integrating Eq.(\ref{eq2}) and using the orthogonality constraint,
we obtain the important property $\sum_{i}^{occ.}C_{xc}^{i\sigma}=0.$

Using Eqs.(\ref{eq1}) and (\ref{eq2}), and dropping spin indices by assuming a paramagnetic
situation, $V_{xc}(z)$ could be formally expressed for real $\xi_{i}(z)$'s
(as is our case) as $V_{xc}(z)=V_{xc\,1}(z)+V_{xc\,2}(z)+V_{xc\,3}(z),$
where\begin{eqnarray}
\label{eq5}V_{xc\,1}(z) & = & \sum_{i}^{occ.}\frac{\left[k_{F}^{i}\xi_{i}(z)\right]^{2}}{2\pi n(z)}\left[u_{xc}^{i}(z)+\Delta\overline{V}_{xc}^{i}\right],\\
\label{eq6}V_{xc\,2}(z) & = & \sum_{i}^{occ.}(k_{F}^{i})^{2}S_{i}(z)/\left[2\pi n(z)\right],\\
\label{eq7}V_{xc\,3}(z) & = & 
\sum_{i}^{occ.}\frac{\left[(k_{F}^{i})^{2}\psi_{i}^{\prime}(z)\xi_{i}^{\prime}(z)-C_{xc}^{i}\xi_{i}^{\prime}(z)^{2}\right]}{2\pi n(z)},
\end{eqnarray}
 with primes denoting derivation with respect to $z.$ It can be shown
that if the shifts in $V_{xc}(z)$ are forced to be zero, the only
term left is $V_{xc\,1}(z).$ This truncated expression
for $V_{xc}(z)$ is identical to the one obtained in Ref.\cite{kli}
for closed systems (KLI approximation). Accordingly, we identify here
that $V_{xc}^{\text{KLI}}(z)\equiv V_{xc\,1}(z).$ In the
$1S$ regime $V_{xc\,2}(z)=V_{xc\,3}(z)\equiv0,$\cite{reboredo03}
then $V_{xc}^{(1S)}(z)=V_{xc}^{\text{KLI}}(z)$. A KLI calculation in the
many-subband situation, requires in our open case to find a self-consistent
solution of Eq. (\ref{eq1}) with $V_{xc}(z)=V_{xc\,1}(z)$, constrained by
the exact condition $\sum_{i}^{occ.}C_{xc}^{i}=0$. This procedure
\textit{univocally} determines $V_{xc1}(z)$. It is important to realize however, that
what we call KLI approximation is not the same as previously considered \cite{kli}. While the
expressions for $V_{xc}^{\text{KLI}}$ are identical in both cases, the boundary conditions are 
different. In our case, we use $\sum_i^{occ}C_{xc}^i=0$. In previous studies, the physical boundary condition
$\overline{V}_{xc}^{m\sigma}=\overline{u}_{xc}^{m\sigma}$ was imposed in order to satisfy the correct asymptotic
behavior ($m$ denoting
the index of the highest-occupied subband).
All results given until
this point include both exchange and correlation. In the $x$-only
version of the OEP method, $V_{xc}\rightarrow V_{x}$.\cite{sahni,dellasala,kummel,kli}

We plot in Fig. \ref{fig1} the $x$-only OEP for several fillings. It is noticeable
the large difference for $V_{x}(z)$ as calculated for $\Delta=\mu-\varepsilon_{1}\rightarrow0^{-}$
(A) or $\Delta\rightarrow0^{+}$ (B), as well as the ``hump''
that $V_{x}(z)$ develops for incipient population of the second subband,
and the fast decay of this ``hump'' after macroscopic occupation
of the band (C).

\begin{figure}
\onefigure[  bb=1cm 8.4cm 19.7cm 26.7cm,
  clip,
  scale=0.42]{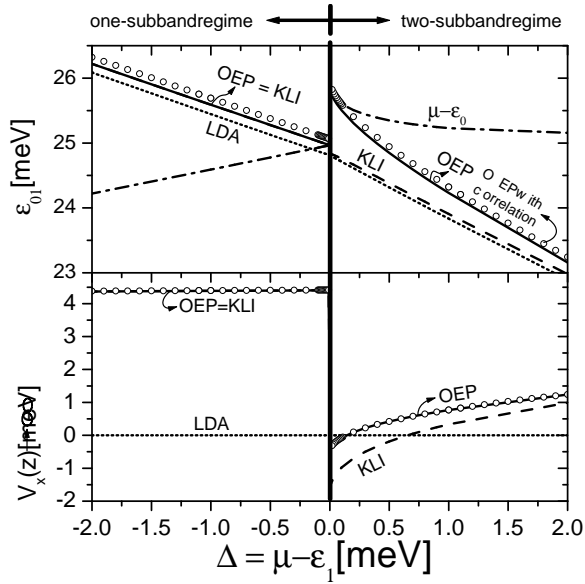}
\caption{Exchange-only energy spectra by different methods. Top panel: inter-subband
energy spacing $\varepsilon_{01}$ and $\mu-\varepsilon_{0}.$ Lower
panel: asymptotic value of the exchange potential $V_{x}(z\rightarrow\infty)=\Delta\overline{V}_{x}^{m}$
(see Eq. (\ref{eq8})). The KLI approximation coincides with the exact result
(OEP) in the $1S$ regime. Open circles: same as above, but including correlation \textit{a la} LDA.}
\label{fig2}
\end{figure}

The $x$-only eigenvalue spectra, in the neighborhood of the $1S$
$\rightarrow$ $2S$ transition, is shown in Fig. \ref{fig2}. The top panel
corresponds to the inter-subband energy $\varepsilon_{01}\equiv\varepsilon_{1}-\varepsilon_{0}$
calculated in three different ways: OEP, KLI and LDA. The lower panel
displays the asymptotic values of the $x$-only potential. We notice
an abrupt \textit{increase} of $\varepsilon_{01}^{\text{OEP}}$ when
the system passes from the $1S$ to the $2S$ regime. This should
be contrasted with the abrupt \textit{decrease} of $\varepsilon_{01}^{\text{KLI}}$,
and the \textit{continuity} of $\varepsilon_{01}^{\text{LDA}}.$ Also
noticeable is the abrupt jump of $V_{x}(z\rightarrow\infty)$ in OEP
and KLI at $\Delta=0$.

In what follows we will provide a brief explanation for each of the
striking results shown in previous figures. We start with the asymptotic
behavior of $V_{x}(z),$ for the many-subband case $m>1$. Following the analysis
of Ref.\cite{kreibich}, and based on the fact that $\xi_{i}(z\rightarrow\infty)\rightarrow e^{-\beta_{i}\, z}$
(disregarding factors involving powers of $z$) with $\beta_{i}>0$
and $\beta_{i}<\beta_{j}$ if $\varepsilon_{i}>\varepsilon_{j},$
one can derive from the asymptotic limit of the differential equation
for the shifts that $\psi_{i<m}(z\rightarrow\infty)\rightarrow e^{-\beta_{m}\, z}.$
From the asymptotic analysis of Eq.(\ref{eq2}), we obtain $\psi_{m}(z\rightarrow\infty)\rightarrow\left[C_{x}^{m}/(k_{F}^{m})^{2}\right]e^{-\beta_{m}\, z}.$
Consequently, \textit{all} the shifts corresponding to occupied subbands
decay exponentially at the same rate. Inserting this result in the
asymptotic expression for $V_{x}(z)$, which amounts to restrict the
sum over the occupied subbands to the last one $(i=m),$ it can be
checked that $V_{x\,2}(z)$ and $V_{x\,3}(z)$ vanish to leading order,
while the contribution from $V_{x\,1}(z)$ remains finite,
\begin{equation}\label{eq8}
V_{x}(z\rightarrow\infty)\rightarrow V_{x\,1}(z\rightarrow\infty)\rightarrow u_{x}^{m}(z\rightarrow\infty)+\Delta\overline{V}_{x}^{m},
\end{equation}
where $u_{x}^{m}(z\rightarrow\infty)\rightarrow-1/z.$\cite{note}
Eq.(\ref{eq8}) coincides with the asymptotic result found in standard
applications of the OEP method to closed systems.\cite{grabo} In such
a case, as discussed above, $V_{x}(z)$ is determined up to an additive constant,
which is fixed arbitrarily choosing $\Delta\overline{V}_{x}^{m}=0$. In
our open configuration, however, the constant is automatically
generated by the self-consistent solution of
Eqs.(\ref{eq1})-(\ref{eq2}). Remembering that
$\Delta\overline{V}_{x}^{m}=\overline{V}_{x}^{m}-\overline{u}_{x}^{m},$
we have no reason to expect a smooth change in the value of the
asymptotic constant at the $1S\rightarrow2S$ transition. This explains
the abrupt jump at $\Delta=0$ in Fig. \ref{fig2}. The KLI
approximation preserves this jump in $\Delta\overline{V}_{x}^{m}$, but
fails quantitatively, giving a jump a 23 \% larger than the exact one.
$V_{x}^{\text{LDA}}(z\rightarrow\infty)=0$, independently of subband
filling.  It is worth at this point to address the important
difference with the related results of Refs.\cite{dellasala,kummel}.
These works are concerned with the asymptotic behavior of the $x$-only
OEP $V_{x}({\mathbf{r}})$ as applied to \textit{closed} systems such
as isolated molecules and metallic clusters. Their main finding is
that $V_{x}({\mathbf{r}})$ can approach different asymptotic values
depending on the spatial direction with which the asymptotic region is
reached. In contrast, the solution of this \textit{open} system 
shows abrupt changes in the asymptotic values of the $x$-only OEP resulting 
from changes in the subband filling, at a \textit{fixed} spatial direction
($z$). 

Let us concentrate now on the the ``hump'' effect observed in
Fig. \ref{fig1}. This effect is due to the contribution $V_{xc\,2}(z)$ in
$V_{xc}(z)$. To see this, let us analyze $V_{xc\,2}(z)$ in the limit
$k_{F}^{1}\rightarrow0^{+}.$ In this almost depleted $2S$ regime,
we can approximate $V_{xc\,2}(z)\simeq C_{xc}^{0}-(k_{F}^{0})^{2}\psi_{0}(z)/\xi_{0}(z).$
Using $S_{0}(z)+S_{1}(z)=0,$ and $C_{xc}^{0}+C_{xc}^{1}=0,$ we find
that $\psi_{0}(z)\simeq C_{xc}^{0}\left[\xi_{0}(z)^{2}-\xi_{1}(z)^{2}\right]/\xi_{0}(z)(k_{F}^{0})^{2}.$
Replacing $\psi_{0}(z)$ in the latter expression for $V_{xc\,2}(z)$,
we obtain $V_{xc\,2}(z)\simeq C_{xc}^{0}\left[\xi_{1}(z)/\xi_{0}(z)\right]^{2}.$
We plot this approximation for $V_{x\,2}(z)$ in Fig. \ref{fig1} (thick line),
showing that it is the ``limiting'' value of the full $V_{x}(z)$
if $\Delta\rightarrow0^{+}$ in the region where the ratio $\xi_{1}(z)/\xi_{0}(z)$
is large, excluding the asymptotic region, which is dominated by $V_{x1}(z)$. Since 
$V_{xc2}$ (and $V_{xc3}$) are neglected in the KLI approximation (as defined above),
the ``hump'' effect is also lost. From this point of view, our KLI is not
a good approximation for slight occupancies of excited subbands.

\begin{figure}
\onefigure[  bb=2.1cm 12.5cm 18.2cm 26.5cm,
  clip,
  scale=0.5]{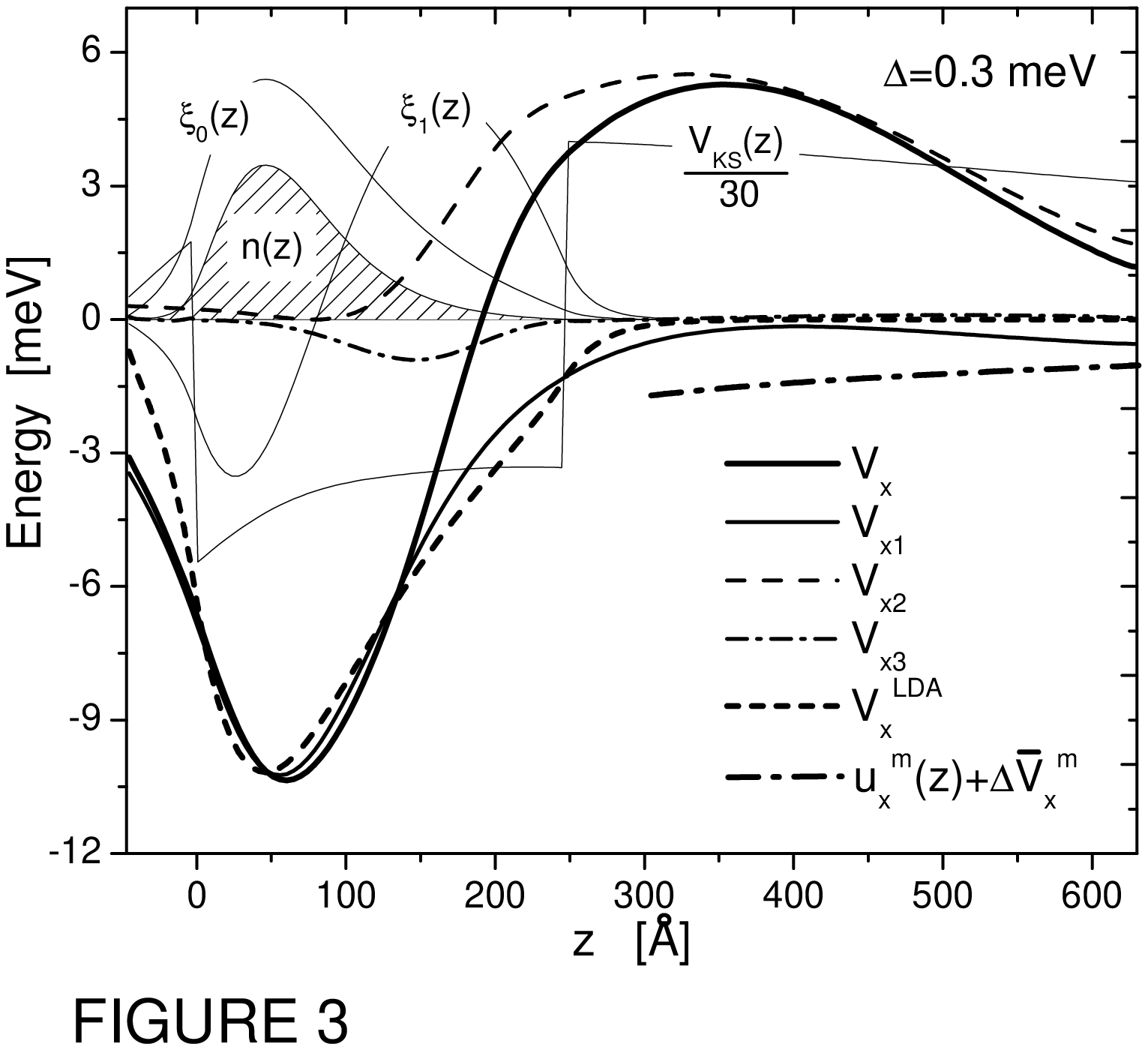}

\caption{
  OEP and LDA $x$-only potentials for $\Delta=0.3$ meV. $\xi_{0}(z)$
  and $\xi_{1}(z)$ are the ground and first-excited subband
  wave-functions, and $n(z)$ is proportional to the density.  The
  asymptotic expression of \(V_{x}(z)\sim
  u_{x}^{m}(z)+\Delta\overline{V}_{x}^{m}\) (Eq.(8)) is only plotted
  in the barrier region.  $V_{x}$ (full line) and $V_{x}$ including
  correlation effects \textit{a la} LDA (not shown) are
  indistinguishable on the scale of the figure.}
\label{fig3}
\end{figure}

To further clarify the role played by the different contributions to
the $x$-only OEP, we display them in Fig. \ref{fig3}. It is clear from
this figure that the ``hump'' effect is due to $V_{x\,2}(z),$ while
$V_{x\,1}(z)$ dominates in the high-density region and gives also the
main contribution to \(V_x(z)\) in the asymptotic region, according to
the result of Eq.(\ref{eq8}).  The contribution from $V_{x\,3}(z)$ has
been found to be small in all the analyzed situations.

The building of this significant barrier explains, in turn, the abrupt
increase of $\varepsilon_{01}^{\text{OEP}}$under second subband occupation:
as the barrier lies about the same zone where the weight of $\xi_{1}(z)$
is concentrated, it affects mainly the electrons in the second subband,
resulting in an increase of the inter-subband spacing $\varepsilon_{01}^{\text{OEP}}$.
However this increase is the result of two opposite contributions:
the barrier building, which tends to increase $\varepsilon_{01}^{\text{OEP}},$
and the abrupt jump in the value of $V_{x}(z\rightarrow\infty)$,
which tends to decrease $\varepsilon_{01}^{\text{OEP}}.$\cite{aceffect}
The fact that the sum of the two contributions result in a net increase
of $\varepsilon_{01}^{\text{OEP}},$ implies that the repulsive barrier
effect overcomes the decrease of $V_{x}(z\rightarrow\infty)$. This
latter effect is the only one included in $\varepsilon_{01}^{\text{KLI}},$
that leads to an abrupt decrease of $\varepsilon_{01}^{\text{KLI}}$
at the $1S$ $\rightarrow$ $2S$ transition. Finally $\varepsilon_{01}^{\text{LDA}},$
lacking from the {}``hump'' and asymptotic value effects, displays
a smooth transition at $\Delta=0$. Several works in the past have
addressed the (surprising) issue of the discontinuity of the exact
$V_{xc}({\mathbf{r}})$ as a function of the electron number.\cite{kli,discont,leeuwen,lannoo}
Such a $V_{xc},$ for a filled-subshell system with $N$ electrons,
must essentially jump by a constant for finite ${\mathbf{r}}$ as
the number of electrons is allowed to vary from $N-\delta$ to $N+\delta$
where $\delta$ is a positive infinitesimal. This non-analytic behavior
of $V_{xc}$ leads in turn to discontinuities on the $N$-dependence of
the eigenvalues, which are reminiscent of the results presented in Fig. 2. 
In our case, $V_{x}$ can also in principle change discontinuously by a constant at the 
$1S\rightarrow2S$ transition. Instead, our calculation predicts that the system 
shows another type of discontinuous behavior, replacing $V_x(z)$ (curve A in Fig. 1)
by \textit{another} function $V_x(z)$ (curve B in Fig. 1). The resulting $V_x(z)$ induces
a charge  transfer from the minority subband to the 
other optimizing  the exchange energy by maximizing overlaps. 
A constant shift in the exchange potential does not favor such transfer. 
 We have assumed that exchange is the dominant contribution although 
correlations could  contribute subband discontinuities \cite{lannoo} . 
In order to \textit{estimate} the importance of correlation 
on the \textit{x}-only results, and considering the good agreement between
$V_x$ and $V_x^{\text{LDA}}$ in the high density region inside the quantum well 
displayed in Fig. 3, we show in Fig. 2 with open circles the effect of 
replacing \(V_{x}^{\text{OEP}}(z)\) by \(V_{x}^{\text{OEP}}(z)+V_{c}^{\text{LDA}}(z)\), both 
on \(\varepsilon^{\text{OEP}}_{01}\) and \(V_{x}^{\text{OEP}}(z\rightarrow \infty)\). This 
amounts to re-calculate the self-consistent solutions of Eqs.(\ref{eq1}) and (\ref{eq2}) for 
all \(\Delta\), with \(V_{x}^{\text{OEP}}(z)\) obtained according to the OEP procedure, while 
\(V_c^{\text{LDA}}\equiv \delta E_c^{\text{LDA}}/\delta n(z)\).\cite{rigamonti} The correction 
on \(\varepsilon^{\text{OEP}}_{01}\) is extremely small (\(\sim 0.1\) meV for all values of 
\(\Delta\)), and could be understood as resulting from a shift of the \textit{x}-only value 
of \(\varepsilon_{01}\) under a small (perturbative) correlation potential. The impact of 
LDA correlation on the discontinuity at \(\Delta =0\) in \(\varepsilon^{\text{OEP}}_{01}\) and 
\(V_{x}^{\text{OEP}}(z\rightarrow \infty)\) is negligible.

Let us discuss the results for this system, in the more general context
of the semiconductor gap problem.\cite{lannoo} Note that if the gap is defined
as the energy difference of exchanging an electron or hole between
a given system and a reservoir\cite{kurth,grabo} then, $i$) it is
necessary to consider an open system, and $ii$) the gap will be affected
by a sudden change of $V_{x}(\mathbf{r})$ under an infinitesimal
occupation of the conduction band as observed in this work. This study
also suggest that going beyond KLI might be required to describe exchange
contributions to the gap. The strong decrease of the barrier-like
component on $V_{x}(z)$ in Fig. 1 
for $\Delta\simeq0.01\,\text{meV}\rightarrow0.3\,\text{meV}$
implies a significant reduction of the exchange contribution to $\varepsilon_{01}^{\text{OEP}}$.
This reduction of $\varepsilon_{01}^{\text{OEP}}$ resembles the one
observed in the gap of semiconductors under intense photo-excitation
(band-gap renormalization). The $\varepsilon_{01}^{\text{OEP}}$ results
of Fig. 2 for $\Delta\gtrsim0$ suggest an exchange contribution to
the band-gap renormalization of the same order of magnitude than the
$\Delta=0$ gap discontinuity itself. This should be contrasted with the smaller reduction 
rate with increasing density displayed by 
$\varepsilon_{01}^{\text{KLI}}$ and $\varepsilon_{01}^{\text{LDA}}$.

In summary, we have shown that the $x$-only OEP has a number of interesting
properties when applied to an open 2DEG: \textit{i)} under depopulation
of the highest-occupied subband, $V_{x}(z)$ develops a barrier-like
structure in the spatial region where the density starts to be dominated
by electrons in that subband; \textit{ii)} the size of the barrier
increases dramatically when the highest subband occupation is tuned
to complete depletion; \textit{iii)} $V_{x}(z\rightarrow\infty)$
strongly depends on subband filling, showing an abrupt jump at the
$1S$ $\rightarrow$ $2S$ transition; and \textit{iv)} as a consequence
of the combined effect of \textit{ii)} and \textit{iii)}, the inter-subband
energy is discontinuous at the $1S$ $\rightarrow$ $2S$ transition.
Correlation effects,
  estimated \textit{a la} LDA, are found to be very small for this system,
  giving hope for its rigorous and systematic consideration following
  the lines of the KS perturbative approach.\cite{gorling} The results for the present system offer insights into fundamental
problems related to the band-gap in semiconductors.

This work was partially supported by CONICET under grant PIP 2753/00.
S. R. acknowledges financial support from CNEA-CONICET. C.R.P. is
a fellow of CONICET. F.A.R. acknowledges the auspices of the U.S.
Department of Energy at the University of California/Lawrence Livermore
National Laboratory under contract no. W-7405-Eng-48.


\begin{thebibliography}{0}

\bibitem{parr}  
\Name{Parr R. G. \and Yang W.} 
\Book{Density Functional Theory of Atoms and Molecules} 
\Publ{Oxford University Press, New York}
\Year{1989}.

\bibitem{kurth}  
\Name{ Perdew J. P. \and  Kurth S.} 
\Book{A Primer in Density Functional Theory}, 
\Editor{Fiolhais C., Nogueira F. \and Marques M.}
\Publ{Lecture Notes in Physics, Springer, Berlin} 
\Year{2003}.

\bibitem{grabo}  
\Name{Grabo T. \textit{et al.}} 
\Book{Strong Coulomb Interactions
in Electronic Structure Calculations: Beyond the Local Density Approximation}
\Editor{ Anisimov V. I.}
\Publ{Gordon and Breach, Amsterdam}
\Year{2000}.

\bibitem{sahni}  \Name{ Sahni V.,  Gruenebaum J.\and  Perdew J. P.}
\REVIEW{Phys. Rev. B}{26}{1982}{4371}.
  
\bibitem{zunger}  \Name{Perdew J. P.\and Zunger A.} 
\REVIEW{Phys. Rev. B }{23}{1981}{5048}.

\bibitem{dellasala}  
\Name{Della Sala F.\and G\"{o}rling A.} 
\REVIEW{Phys. Rev. Lett.}{89}{2002}{33003}; 
{\textit{ibid}}, \REVIEW{J. Chem. Phys.}{116}{2002}{5374}.

\bibitem{semi}  
\Name{Bylander D. M.\and  Kleinman L.} 
\REVIEW{Phys. Rev. Lett.}{74}{1995}{3660};
\Name{St\"{a}dele M.  \textit{et al.}}
\REVIEW{Phys. Rev. Lett.}{79}{1997}{2089};
\Name{ Kim Y.-H. \and  G\"{o}rling A.} 
\REVIEW{Phys. Rev. Lett.}{89}{2002}{96402}.

\bibitem{van}  
\Name{ van Gisbergen S. J. A. \textit{et al.}} 
\REVIEW{Phys. Rev. Lett.}{83}{1999}{694}.

\bibitem{rigamonti}  
\Name{ Rigamonti S., Reboredo  F. A. \and  Proetto C. R.}
\REVIEW{ Phys. Rev. B}{68}{2003}{235309}.


\bibitem{method}We have used the method of Ref. \cite{kummel} for 
the solution of this differential equation, and followed their numerical strategy for 
the self-consistent solution of Eqs.(\ref{eq1}) and (\ref{eq2}).

\bibitem{kummel}  
\Name{Kummel S.\and Perdew J. P.} 
\REVIEW{Phys. Rev. Lett.}{90}{2003}{043004}
{\textit{ibid}}, \REVIEW{Phys. Rev. B}{68}{2003}{035103}.

\bibitem{kli}  
\Name{ Krieger J. B., Li Y. \and  Iafrate G. J.} 
\REVIEW{Phys. Rev. A}{45}{1992}{101}.

\bibitem{reboredo03} 
\Name{ Reboredo F. A. \and  Proetto C. R.} 
\REVIEW{Phys Rev. B}{67}{2003}{115325}.

\bibitem{kreibich}  
\Name{Kreibich T.  \textit{et al.}}
\REVIEW{Adv. Quantum Chem.}{33}{1998}{31}.

\bibitem{note}  Note that Eq.(\ref{eq8}) is exactly the result that we have obtained
for {\textit{all}} \(z\) in the \(1S\) regime (\(m=0\)).

\bibitem{aceffect} Note that the potential bottom remains nearly constant about \(-10\) meV in Fig. \ref{fig1}. As \( V_{x}^{A}(z \rightarrow \infty )>V_{x}^{B}(z \rightarrow \infty ) \), this means that the potential is deeper if \( \Delta\rightarrow 0^{-} \)(case A), as compared with \( \Delta \rightarrow 0^{+} \) (case B).


\bibitem{discont}  
\Name{ Perdew J. P.{\textit{et al.}}}
\REVIEW{Phys. Rev. Lett.}{49}{1982}{1691};
\Name{Perdew J. P.  \and  Levy M.} 
\SAME{51}{1983}{1884}
\Name{Schl\"{u}ter M. \and  Sham L. J.} 
\SAME{51}{1983}{1888}.

\bibitem{leeuwen}  
\Name{ van Leeuwen R.,   Gritsenko O. V. \and  Baerends E. J.} 
\REVIEW{Z. Phys. D}{33}{1995}{229}.

\bibitem{lannoo}
\Name{Lannoo M., Schl\"{u}ter M. \and Sham L.J.}
\REVIEW{Phys. Rev. B}{32}{1985}{3890};
\Name{Godby R.W., Schl\"{u}ter M.\and Sham L.J.}
\SAME{37}{10159}{1988}.

\bibitem{gorling}
\Name{G\"{o}rling A. \and Levy M.}
\REVIEW{Phys. Rev. B}{47}{1993}{13105}.

\end{thebibliography}
\end{document}